\documentclass[twocolumn,pra,showpacs,floatfix,amsmath,amsfonts]{revtex4}
\usepackage[latin9]{inputenc}
\usepackage{xcolor}
\usepackage{amsmath}
\usepackage{amssymb}
\usepackage{graphicx}
\usepackage[unicode=true, bookmarks=false, breaklinks=false,pdfborder={0 0 1},backref=section,colorlinks=true] {hyperref}
\hypersetup{ linkcolor=blue,citecolor=blue}
\makeatletter
\@ifundefined{textcolor}{}
{%
 \definecolor{BLACK}{gray}{0}
 \definecolor{WHITE}{gray}{1}
 \definecolor{RED}{rgb}{1,0,0}
 \definecolor{GREEN}{rgb}{0,1,0}
 \definecolor{BLUE}{rgb}{0,0,1}
 \definecolor{CYAN}{cmyk}{1,0,0,0}
 \definecolor{MAGENTA}{cmyk}{0,1,0,0}
 \definecolor{YELLOW}{cmyk}{0,0,1,0}
}

\usepackage{epstopdf}

\@ifundefined{textcolor}{}{%
\definecolor{BLACK}{gray}{0}
\definecolor{WHITE}{gray}{1}
 \definecolor{RED}{rgb}{1,0,0}
 \definecolor{GREEN}{rgb}{0, 1, 0}
\definecolor{BLUE}{rgb}{0,0,1}
\definecolor{CYAN}{cmyk}{1,0,0,0}
\definecolor{MAGENTA}{cmyk}{0,1,0,0}
\definecolor{YELLOW}{cmyk}{0,0,1,0}
}
\usepackage{graphics}\usepackage{epsfig}
\allowdisplaybreaks

\makeatother

\begin{document}
\title{Compacton existence and spin-orbit density dependence in Bose-Einstein condensates}
\author{F.Kh. Abdullaev$^{1}$, M. S. A. Hadi$^{2}$, B. Umarov$^{2}$ and
L. A. Taib$^{2}$, and Mario Salerno$^{3}$}
\affiliation{$^{1}$ Physical-Technical Institute, Uzbekistan Academy of Sciences,
Tashkent, Bodomzor yuli, 2-b, Uzbekistan}
\affiliation{$^{2}$ Department of Computational and Theoretical Sciences,, Kulliyyah
of Science, International Islamic University Malaysia, 25200 Kuantan,
Pahang, Malaysia}
\affiliation{$^{3}$ Dipartimento di Fisica ``E.R. Caianiello'', CNISM and INFN
- Gruppo Collegato di Salerno, Universit� di Salerno, Via Giovanni
Paolo II, 84084 Fisciano (SA), Italy}
\date{\today}
\begin{abstract}
We demonstrate the existence of compactons matter waves in binary
mixtures of Bose-Einstein condensates (BEC) trapped in deep optical
lattices (OL) subjected to equal contributions of intra-species Rashba
and Dresselhaus spin-orbit coupling (SOC) under periodic time modulations
of the intra-species scattering length. We show that
these modulations lead to a rescaling of the SOC parameters
that involves the density imbalance of the two components. This gives rise to a density dependent
SOC parameters that strongly influence the existence and the stability of compacton matter waves.
The stability of SOC-compactons is investigated both by linear stability
analysis and by time integrations of the coupled  Gross-Pitaevskii
equations. We find that SOC restricts the parameter ranges for stable
stationary SOC-compacton existence but, on the other side, it gives
a more stringent signature of their occurrence. In particular, SOC-compactons
should appear when the intra-species interactions and the number of
atoms in the two components are perfectly balanced (or close to be
balanced for metastable case). The possibility to use SOC-compactons
as a tool for indirect measurements of the number of atoms and/or
the intra-species interactions, is also suggested.
\end{abstract}
\maketitle

\section{Introduction}

Recently theoretical and experimental investigations of Bose-Einstein
condensates (BEC) based on Floquet engineering (FE) of the different
parameters have been performed (see for example the review~\cite{Bukov}).
The FE usually involves linear optical lattices, periodically shaken
in time either through frequency or amplitude~\cite{Lin,Galitski},
or nonlinear optical lattices in which FE is realized through the
modulation of the interactions~\cite{Abdullaev1}. Periodic modulations
of the scattering lengths can also be used to emulate synthetic dimensions
and to create density-dependent gauge fields \cite{Greschner}, this
being a research area of rapidly growing interest connected with interesting
physical phenomena, including pair superfluidity, exactly defect-free
Mott insulator states~\cite{Rapp}, etc. In the case of BEC with modulated
interactions in double-well potential, the investigation shows that
the tunneling transition amplitude between two wells depends on the
relative imbalance of the atomic population between wells. This phenomenon
leads to the suppression of the tunneling between wells for specific
values of the imbalance~\cite{Gong}, a prediction that was recently
experimentally confirmed in~\cite{Meinert}.

For Bose-Einstein condensates trapped in a deep optical lattice (OL)
the phenomenon of tunneling suppression leads to the existence of
a new form the localized matter waves, also called BEC compactons,
in which the density is strictly localized on a compact domain without
any exponential decay tail at the boundary~\cite{Abdullaev1}. BEC
compactons can exist not only in ordinary (e.g. single component)
BEC but also in binary BEC mixtures~\cite{Abdullaev2,Abdullaev3}
as well as in multidimensional contexts~\cite{multidim}. In the case
of a deep modulation of the potential, the problem can be reduced
to the nonlinear lattice described by a discrete nonlinear Schr\"odinger
(DNLS) model with modulated nonlinearity. In this case the maximum
localization is achieved with one-site compactons, i.e., excitations
in which all the matter remains localized in a single well of the
OL.

In binary BEC mixtures, however, another type of coupling, besides
the one induced by the two-body inter- and intra-species interactions,
is also possible, namely the spin-orbit--coupling (SOC). BEC with
spin-orbit coupling (SOC-BEC) is presently receiving a lot of attention
as with the ultra-cold atoms variety of synthetic SOC can be engineered
and controlled by external laser fields. The experimental realization
of SOC-BEC for the case of binary mixtures was reported in \cite{Lin,Galitski},
while nonlinear excitations of intra-SOC and inter-SOC, i.e., with
the spatial derivative of the SOC term acting inside each species
or between the species, respectively, were theoretically investigated
in \cite{Belicev}. For SOC-BEC trapped in deep OL, the existence
of strongly localized excitation in the form of discrete breathers
was theoretically investigated in \cite{Salerno1} and the SOC tunability
induced by periodic time modulations of the Zeeman term was demonstrated
in \cite{Salerno2}.

Very recently the nonlinearity management of the spin-orbit coupled
BEC has been investigated in \cite{SM} where it has been shown that
for slow time periodic modulations of the nonlinearity, resonances
between the frequencies of the modulation and of the intrinsic nonlinear
modes (solitons) can lead to the appearance of instability and stability
tongues for the solitons.

We remark that only a few studies exist for the case of SOC-BEC in deep
OL subjected to nonlinear management \cite{Lin}. This case is rather
interesting because, compared to previous studies, the presence of
the SOC term may affect the compacton matter waves excitations found
in absence of SOC \cite{Abdullaev1,Abdullaev2,Abdullaev3}. In particular,
the spatial derivative of the SOC term may interfere with the tunneling
of atoms between adjacent wells, making it unclear whether SOC-compactons
may still be possible. At the best of our knowledge, this problem is
completely uninvestigated.

The aim of the present paper is to provide an answer to the above
problem, i.e., we investigate the formation of binary matter-wave
compactons in the presence of SOC. For this, we consider an intra-SOC-BEC
mixture in a deep OL subjected to strong nonlinear management (SNM) consisting
in  time-periodic modulations of the inter-species scattering length.
Using the averaging method, we show that the SNM induces a rescaling
of the inter-well tunneling constant and of the SOC parameters that
depend on the density imbalance between neighboring
sites, this making the conditions for the compacton existence and
stability much more involved.

In particular, we show that the interplay
between SOC and inter-well tunneling produce a big effect on the stability
of the compactons which appears very much reduced by the presence
of SOC. This is true even for the case of a single site compacton
for which the existence range is completely independent on SOC parameters.
A detailed linear stability analysis shows that single site compactons
are stable for a wide range of inter-well tunneling when the SOC parameter
is relatively small. The SOC range, however, can be made wider by
increasing the Rabi frequency. This is true also for compactons localized
on more than one site.

It is worth to note that the possibility to induce by means of  SNM a local density imbalance dependence on the SOC  parameters 
is a fact that  persists independently on compacton existence and, as far as we know,  was not reported before.
Very recently a density-dependent SOC was also introduced in ultracold Bose and Fermi gases in terms of interaction assisted Raman processes \cite{Xu2021}. We remark that our approach
is different from this, being completely independent on indirect Raman processes
(the density dependence of the SOC in our case is only due to
scattering length modulations).

The paper is organized as follows. In Section II, we introduce the
model equations for the system and show in detail the averaging procedure
to derive the effective equations that are valid in the limit of strong
modulations of the interspecies scattering length. In Sections III and IV
we study the SOC-compacton stationary and existence conditions, respectively. In Section V we present numerical results for the cases of one, two and  three sites compactons and investigate their stability both by linear analysis and by direct time integrations of the equations of motion.
Finally, in Section VI  the main results of this paper are briefly summarized.

\section{Model equations and averaging}

As a model for a quasi-one-dimensional BEC mixture trapped in an OL
in the presence of SOC, we consider the following coupled Gross-Pitaevskii
equations (CGPE)
\begin{eqnarray}
i\hbar\frac{\partial\Psi}{\partial t} & = & H\Psi\equiv\left[H_{0}+H_{nl}\right]\Psi,\label{eq:001}\\
H_{0} & \equiv & \frac{p_{x}^{2}}{2m}+\frac{\hbar\kappa}{m}p_{x}\sigma_{z}+V_{ol}(x)+\hbar\bar{\Omega}\sigma_{x},\nonumber \\
H_{nl} & \equiv & 2\hbar\omega\left(\begin{array}{cc}
\sum_{j}a_{1j}\left|\Psi_{j}\right|^{2} & 0\\
0 & \sum_{j}a_{j2}\left|\Psi_{j}\right|^{2}
\end{array}\right),\nonumber
\end{eqnarray}
where $\sigma_{x,z}$ are usual Pauli matrices and $\Psi=\left(\Psi_{1},\Psi_{2}\right)^{T}$
denotes the two-component wave function normalized to the total number
of atoms. This quasi-1D model can be derived from a more general
three-dimensional setting by considering a trapping
potential with the transversal frequency $\omega_\perp$ much larger than
the longitudinal one, $\omega_\perp \gg \omega_\parallel$.
The linear part, $H_{0}$, of the Hamiltonian $H$ includes
the kinetic energy $p_{x}^{2}/(2m)$ of the condensate, the diagonal
SOC term of strength $\kappa$, the off-diagonal Rabi oscillation
term of frequency $\Omega$, and an optical lattice in the $x$-direction
represented by a periodic potential $V_{ol}(x)\sim\sin^{2}\left(k_{L}x\right)$
where $k_{L}$ is the lattice wavenumber. Note that the SOC term is
diagonal in the two components, meaning that we are considering the
intra-SOC case (a similar analysis can be done for the inter-SOC case
with an  off-diagonal $\kappa$ term and a diagonal Zeeman term).
The nonlinear part, $H_{nl}$,
includes contact interactions, with $a_{jj}\,\,(j=1,2)$ and $a_{12}$
the two-body intra-species and inter-species scattering lengths, respectively.
By rescaling variables and component wavefunctions according to
\begin{eqnarray}
x & \rightarrow & \frac{x}{k_{L}},\,\,\,t\rightarrow\omega_{R}t,\,\,\,\mathrm{where}\,\,\omega_{R}\equiv\frac{E_{R}}{\hbar}\equiv\frac{\hbar k_{L}^{2}}{2m};\nonumber \\
V_{ol}(x) & \rightarrow & E_{R}V(x)=E_{R}V_{0}\cos(2x),\label{eq:002}\\
\Psi_{j} & \equiv & \sqrt{\frac{\omega_{R}}{2 \omega_\perp a_{0}}}\psi_{j}(x,t),\nonumber
\end{eqnarray}
the Eq.~\eqref{eq:001} can be put in the dimensionless form:
\begin{eqnarray}
i\frac{\partial\psi_{j}}{\partial t} & = & \left[-\frac{\partial^{2}}{\partial x^{2}}+V\left(x\right)-\left(-\right)^{j}i\frac{2\kappa}{k_{L}}\frac{\partial}{\partial x}\right]\psi_{j}\nonumber \\
 & + & \left(\frac{a_{jj}}{a_{0}}\left|\psi_{j}\right|^{2}+\frac{a_{j,3-j}}{a_{0}}\left|\psi_{3-j}\right|^{2}\right)\psi_{j}\label{eq:003}\\
 & + & \frac{\bar{\Omega}}{\omega_{R}}\psi_{3-j},\;\;\;j=1,2, \nonumber
\end{eqnarray}
with $a_0$ denoting the background scattering length and $E_R$ the lattice recoil energy.

By expanding the two component fields as:  $\psi_1(x,t)= \sum_{n} u_n(t) w(x-na)$,  $\psi_2(x,t)= \sum_n v_n(t) w(x-na)$ \cite{alfimov}, with $w(x-na)$ the lower band Wannier functions of the underlying linear system, one obtains in the tight-binding approximation appropriate for deep optical lattices, the following discrete nonlinear Schr\"odinger system \cite{Salerno1}

\begin{equation}
\begin{aligned}iu_{n,t}= & -\varGamma(u_{n+1}+u_{n-1})+i\sigma(u_{n+1}-u_{n-1})\\
 & +\Omega v_{n}+[\gamma_{1}\left|u_{n}\right|^{2}+\gamma\left|v_{n}\right|^{2}]u_{n},\\
iv_{n,t}= & -\varGamma(v_{n+1}+v_{n-1})-i\sigma(v_{n+1}-v_{n-1})\\
 & +\Omega{u}_{n}+[\gamma\left|u_{n}\right|^{2}+\gamma_{2}\left|v_{n}\right|^{2}]v_{n}.
\end{aligned}
\label{eq:004}
\end{equation}
Here $n$ denotes the lattice site at position $na$, $a$ is the lattice constant, below fixed to $a=1$ without any loss of generality, $\varGamma$ is the inter-site hopping coefficient, $\sigma$
the rescaled SOC strength, $\gamma_{1},\gamma_{2}$ and $\gamma$
are the coefficients of intra and inter-species interactions, respectively. The system \eqref{eq:004}
is of Hamiltonian form with the Hamiltonian given by:
\begin{equation}
\begin{aligned}H= & \sum_{n}\Big[-\varGamma(u_{n+1}u_{n}^{*}+u_{n+1}^{*}u_{n})\\
 & -\varGamma(v_{n+1}v_{n}^{*}+v_{n+1}^{*}v_{n})\\
 & +i\sigma(u_{n+1}u_{n}^{*}-u_{n+1}^{*}u_{n})\\
 & -i\sigma(v_{n+1}v_{n}^{*}-v_{n+1}^{*}v_{n})\\
 & +\Omega(u_{n}v_{n}^{*}+u_{n}^{*}v_{n})\\
 & +\frac{1}{2}(\gamma_{1}\vert u_{n}\vert^{4}+\gamma_{2}\vert v_{n}\vert^{4})+\gamma(t)\vert u_{n}\vert^{2}\vert v_{n}\vert^{2}\Big].
\end{aligned}
\label{eq:005}
\end{equation}
In the absence of nonlinearity, i.e. $\gamma_{1}=\gamma_{2}=\gamma=0$,
the dispersion relation can be obtained by substituting $u_{n}=A\exp\left[i\left(kn-\mu{t}\right)\right]$
and $v_{n}=B\exp\left[i\left(kn-\mu{t}\right)\right]$ into Eq.~\eqref{eq:004},
this giving the following lower and upper branches for the chemical
potentials of the two components:
\begin{equation}
\mu_{\pm}\left(k\right)=-2\varGamma\cos\left(k\right)\pm\sqrt{\Omega^{2}+\sigma^{2}\sin^{2}\left(k\right)}.\label{eq:006}
\end{equation}

In the following we shall investigate the possibility of compacton
formation via an inter-species NM, this being more simple than the
intra-species case since it involves the modulation of just a single
parameter, i.e. the inter-species scattering length. Specifically,
we keep the intra-species nonlinear coefficients, $\gamma_{i},i=1,2$,
in Eq.~\eqref{eq:005} to be constants and assume the inter-species
nonlinearity coefficient $\gamma$ to be a time periodic function
of the form:
\[
\gamma(t)=\gamma^{(0)}+\frac{\gamma^{(1)}}{\epsilon}\cos\left(\frac{\omega t}{\epsilon}\right)
\]
with $\epsilon$ a real parameter which allows to control the strength
of the management and to separate the fast and slow time scales (see
below). Our analysis will be valid in the limit of strong NM which
requires $\epsilon\ll1$. In this case it is possible to eliminate
the explicit time dependence from Eq.~\eqref{eq:005} by means of the
transformation:
\begin{equation}
u_{n}=U_{n}e^{-i\Lambda(\tau)\vert{V}_{n}\vert^{2}},\quad v_{n}=V_{n}e^{-i\Lambda(\tau)\vert{U}_{n}\vert^{2}},\label{eq:007}
\end{equation}
with
\begin{equation}
\Lambda(\tau)=\alpha\sin(\omega\tau),\quad\alpha=\frac{\gamma^{(1)}}{\omega},\quad\tau=\frac{t}{\epsilon},\label{eq:008}
\end{equation}
and to apply the averaging method to eliminate the fast time scale.
The resulting averaged equations for the amplitudes $U_{n},V_{n}$
are given in the Appendix (see Eqs.\eqref{eq:021} and \eqref{eq:022}).

It turns out that also the averaged equations have Hamiltonian structure
(see Appendix) with the same Hamiltonian \eqref{eq:005} of the unmodulated
system but with $u_{n},v_{n}$ replaced by $U_{n},V_{n}$, respectively,
and $\varGamma,\sigma,\Omega$ rescaled according to:
\begin{eqnarray}
 &  & \varGamma\rightarrow\tilde{\varGamma}_{i}\equiv\varGamma J_{0}(\theta_{3-i}^{+}),\nonumber \\
 &  & \sigma\rightarrow\tilde{\sigma}_{i}\equiv\sigma J_{0}(\theta_{3-i}^{+}),\label{eq:009}\\
 &  & \Omega\rightarrow\tilde{\Omega}\equiv\Omega J_{0}(\theta_{0}).\nonumber
\end{eqnarray}
Here $i=1,2$ refer to first and second component respectively, $J_{0}$
is the Bessel function of order zero and $\theta_{1}^{\pm}=\vert U_{n\pm1}\vert^{2}-\vert U_{n}\vert^{2}$,
$\theta_{2}^{\pm}=\vert V_{n\pm1}\vert^{2}-\vert V_{n}\vert^{2}$
and $\theta_{0}=\vert U_{n}\vert^{2}-\vert V_{n}\vert^{2}$. From
this we see that the effect of the modulation is to introduce a dependence
in the parameters $\varGamma,\sigma$ of the unmodulated system, on
the density imbalance between adjacent sites of
the same component, and a dependence on the density imbalance
between the two components on the same site, for $\Omega$.
Obviously the condition for compacton existence, i.e. the complete suppression
of the tunneling at the compacton edges, involves also the SOC parameters
and will have more restrictive conditions to be satisfied.

This can be also intuitively understood by observing that the term proportional to $\sigma_z$ in the Hamiltonian changes the kinetic energy of both components while the term proportional to $\sigma_x$ corresponds to  Rabi-coupling leading to oscillations between the two BEC components. On the other hand, the variation of the kinetic energy due to the $\sigma$ part of the SOC interferes with the natural dispersion of the system, i.e. with the intra-well tunneling $\Gamma$,  while the Rabi term interferes with the stationarity condition of compacton. From this it is clear that the conditions on parameters for the SOC-compacton existence are expected to be more stringent than those in the absence of SOC.

Finally, we remark that while in the density-dependent
SOC approach considered in Ref. \cite{Xu2021} the rescaling of the
parameter $\Omega$ involves both the zeroth and the first-order Bessel
functions, in our case, due to the absence of indirect Raman processes,
only the function $J_{0}$ is involved.

\begin{figure*}[t]
\includegraphics[scale=0.8]{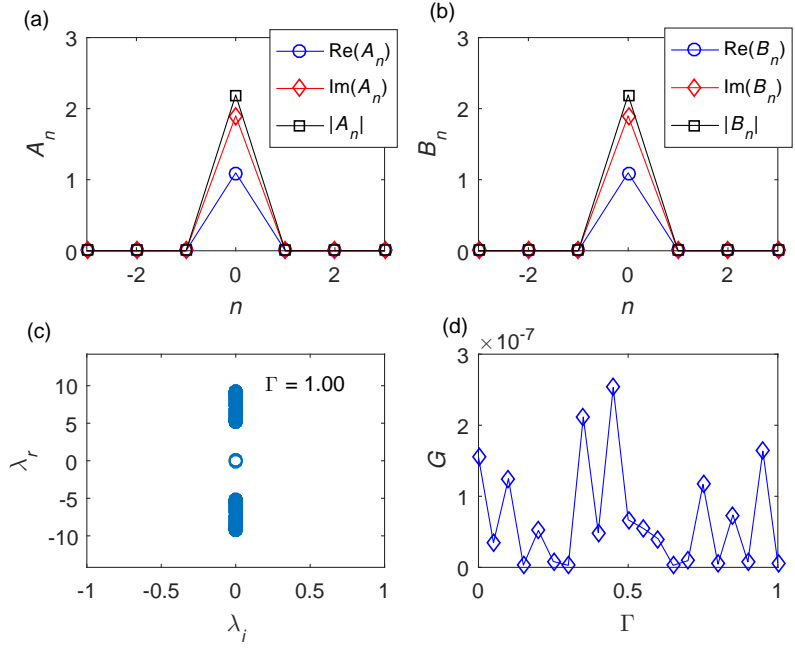} \caption{(a),(b) Binary one-site compactons \eqref{eq:012} localized at $n=0$,
for the case $\sigma=0,\Omega=0$ and with $A_{0}=B_{0}=a+ib$ with
$a=0.5\sqrt{\xi_{1}/\alpha}$ and $b=\sqrt{\xi_{1}/\alpha-a^{2}}$.
Other parameters are fixed as $\varGamma=1,\gamma_{1}=\gamma_{2}=-1$,
$\gamma^{(0)}=-0.5$, $\gamma^{(1)}=0.5$, $\omega=1.0$. (c) The
eigenvalue spectrum at $\varGamma=1$, (d) Gain $G$ versus $\varGamma$.}
\label{Figure01}
\end{figure*}

\section{Stationary conditions for SOC-compactons}

We look for stationary SOC-compactons of the form:
\begin{equation}
U_{n}=A_{n}e^{-i\mu_{1}t},\quad V_{n}=B_{n}e^{-i\mu_{2}t},\label{eq:010}
\end{equation}
with $A_{n}$, $B_{n}$ complex amplitudes and $\mu_{1}$, $\mu_{2}$
chemical potentials for the two BEC components. By substituting these
expressions into the averaged equations for $U_{n}$, $V_{n}$ given
in the Appendix, one finds that all the time dependent exponential
factors can be cancelled out except the ones that are proportional
to $\Omega$, these being of the form:
\begin{eqnarray}
 & \Omega e^{-i(\mu_{1}+\mu_{2})t}\Big\{\alpha A_{n}^{2}B_{n}^{*}e^{2i\mu_{2}t}J_{1}(\alpha\theta_{0})\;\;\;\;\;\;\;\;\;\;\;\;\\
 & \;\;\;\;\;\;\;\;-e^{2i\mu_{1}t}B_{n}\Big[J_{0}(\alpha\theta_{0})-\alpha|A_{n}|^{2}J_{1}
 (\alpha\theta_{0})\Big]\Big\} \nonumber
\label{equaz11}
\end{eqnarray}
for Eq.~\eqref{eq:023}, and similar expression for Eq.~\eqref{eq:024}
but with the exchanges $A\leftrightarrow B$ and $\mu_{1}\leftrightarrow\mu_{2}$.
From these expressions it is clear that stationary SOC-compactons
are possible only if $\Omega=0$ or if the chemical potentials of
the two species are equal, i.e. $\mu_{1}=\mu_{2}\equiv\mu$. The corresponding
steady-state equations then are:
\begin{equation}
\mu_{1}A_{n}=F_{1}|_{\Omega=0},\,\,\mu_{2}B_{n}=F_{2}|_{\Omega=0},\label{eq:011}
\end{equation}
for case $\Omega=0$, and
\begin{equation}
\mu A_{n}=F_{1},\,\,\,\mu B_{n}=F_{2},\label{eq:012}
\end{equation}
for case $\mu_{1}=\mu_{2}=\mu$. Here $F_{1}$ and $F_{2}$ denote
the right hand side of Eqs.\eqref{eq:023} and \eqref{eq:024}, respectively,
but with the replacements: $U_{n}\rightarrow A_{n}$, $V_{n}\rightarrow B_{n}$.
We remark that although the above stationary conditions are specific
of the intra-SOC system, similar conditions would appear also for
the inter-SOC system \cite{Note2}.

As we shall see in the next section, the condition $\Omega=0$ is
too restrictive and can be satisfied only for one-site compactons.
In this case, however, although the stability of the solutions will
be influenced by the SOC, the analytical expression of the solutions
will be the same as for the non-SOC case.

Note that Eq.~\eqref{eq:012} implies the equality of the
number of atoms $N_{1}, N_{2}$ in the two components, this being true in
general also for multi-site SOC-compactons. This can be
easily understood from the fact that for $N_{1}\neq N_{2}$ and
$\Omega\ne0$, there will be oscillations in the number of atoms that
will make the zero tunneling conditions at the compacton edges impossible
to satisfy.  The only way to avoid this is to equilibrate
the two components. This, however, requires the  additional restriction
of the equality of the intra-species interactions (see below).

\section{Existence conditions for SOC-compactons}

The conditions for the existence of stationary compactons are obtained
from Eqs.\eqref{eq:011}, once the following compacton ansatz is adopted
\begin{equation}
\begin{gathered}
A_{n},\,B_{n}\neq0\;\mbox{\;\;\;\;\; if\;\;\ensuremath{n_{0}-s\le n\le n_{0}+s}},\\
A_{n},\,B_{n}=0\;\;\;\;\;\;\mbox{otherwise,\;\;\;\;\;\;\;\;\;\;\;\;\;\;\;\;\;\;\;}
\end{gathered}
\label{eq:013}
\end{equation}
with $n_{0}-s$ and $n_{0}+s$ denoting the left and right edge of
the compacton, respectively, and $s+1$ its width. Since the coupling
in the stationary equations involves only first next neighbors, the
substitution of the ansatz $\eqref{eq:013}$ into either Eq.~\eqref{eq:011}
or Eq.~\eqref{eq:012} will give $2(s+3)$ nontrivial equations
in correspondence of sites $n_{0}-s-1,...,n_{0},...,n_{0}+s+1,$,
with all the others, $n\leq n_{0}-s-2$ and $n\ge n_{0}+s+2$, automatically
satisfied. The compacton existence relies on the possibility to solve
these $2(s+3)$ equations for the nonzero amplitudes by achieving
both the tunneling suppression at the compacton edges and the fixing
of the chemical potentials.

\begin{figure*}
\includegraphics[scale=0.8]{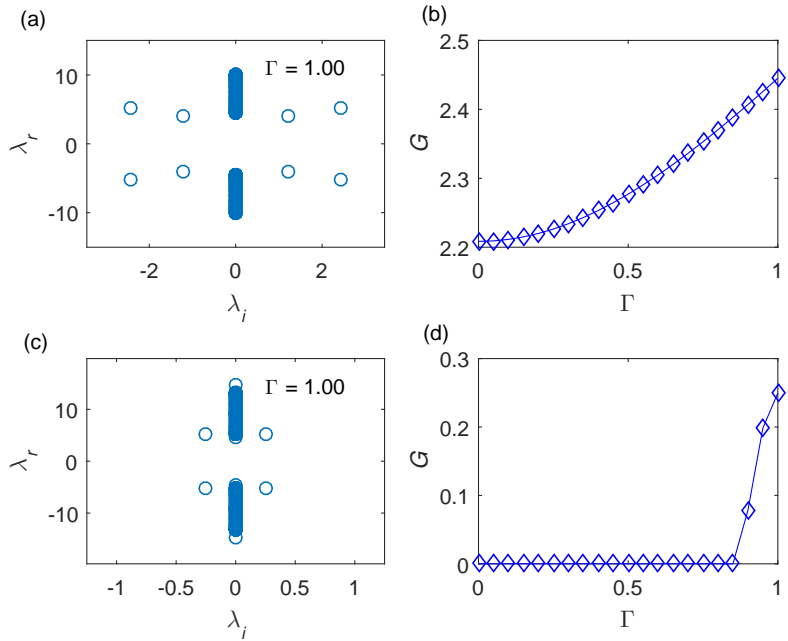} \caption{Parameters for panels (a) and (b) are fixed as in panels (c) and (d)
of Fig. \ref{Figure01} except $\sigma=1$ and for panels (c) and
(d), the parameters are fixed as in (a) and (b) except $\Omega=-2$.}
\label{Figure02}
\end{figure*}

\begin{figure*}[p]
\includegraphics[scale=0.8]{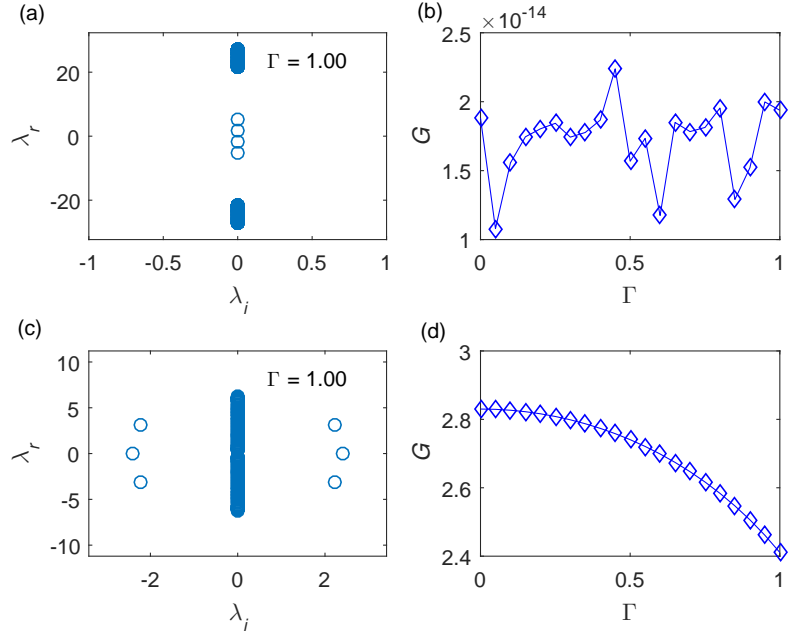} \caption{Linear stability analysis for the one-site compactons , (a) The eigenvalue
spectrum for case $\varGamma=1,\gamma_{1}=\gamma_{2}=-1$, $\gamma^{(0)}=-1$,
$\gamma^{(1)}=0.2$, $\omega=1.0$, $\Omega=-0.4,\sigma=1$, (b) Gain
$G$ versus $\varGamma$ with parameters as in (a), (c) The eigenvalue
spectrum for case $\varGamma=1,\gamma_{1}=\gamma_{2}=-1$, $\gamma^{(0)}=-0.2$,
$\gamma^{(1)}=1$, $\omega=1$ , $\Omega=-0.4,\sigma=1$, (d) Gain
$G$ versus $\varGamma$ with parameters as in (c).}
\label{Figure03}
\end{figure*}

\begin{figure*}[p]
\includegraphics[scale=0.8]{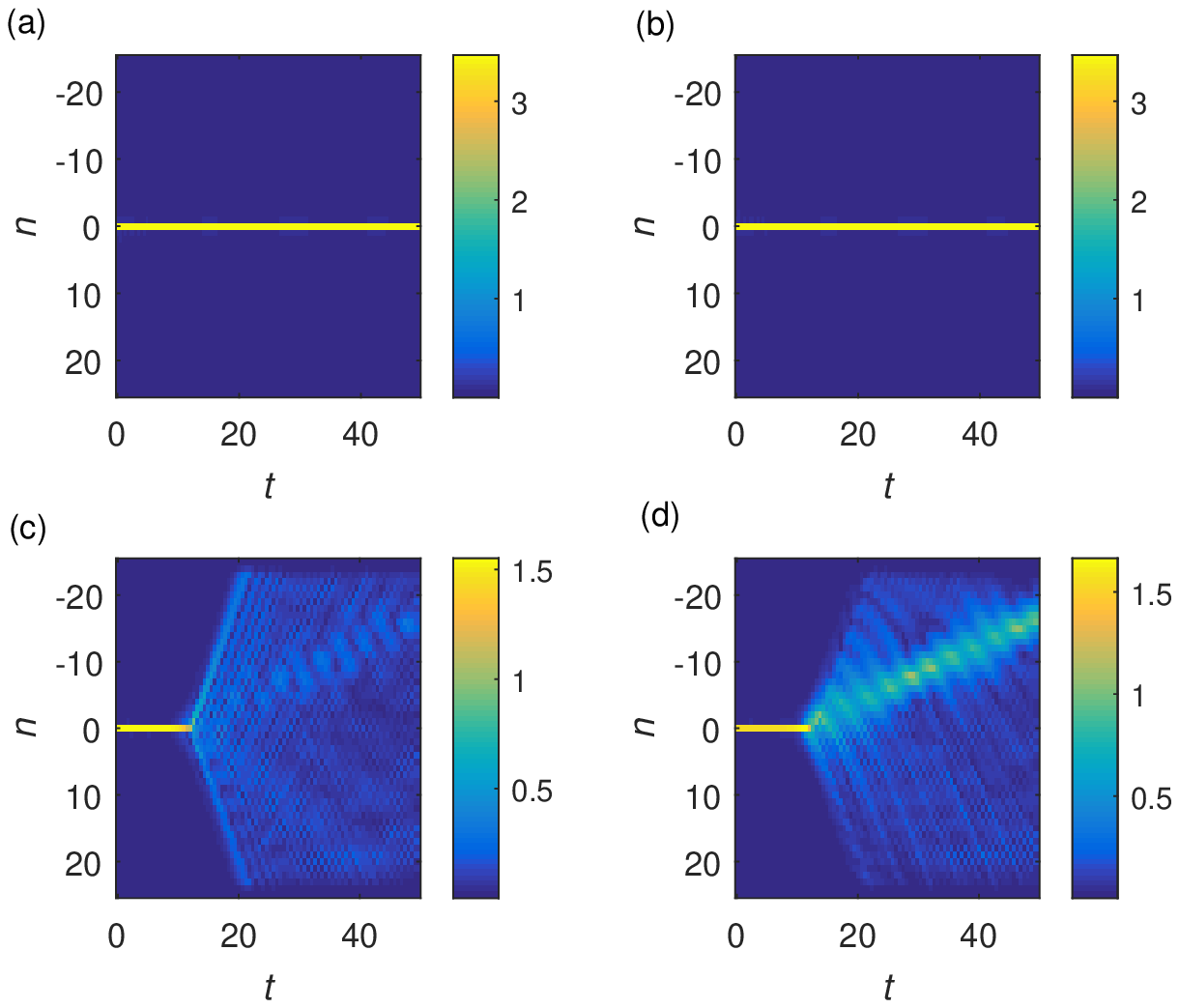}

\caption{Space-time evolution of the on-site compactons by solving Eq.~\eqref{eq:004}
with $\epsilon=0.01$ , (a) $\left|u_{n}\right|$ and (b) $\left|v_{n}\right|$
for stable case as in Fig.\ref{Figure03}(a) while (c) $\left|u_{n}\right|$
and (d) $\left|v_{n}\right|$ for unstable case as in Fig. \ref{Figure03}(c).}

\label{Figure04}
\end{figure*}

\subsection{One-site SOC-compactons}

For a one-site compacton the condition $\Omega=0$ implies existence
conditions completely independent on the SOC parameters $\sigma,\Omega$
and therefore will be similar to the case of BEC mixtures in absence
of SOC considered in previous papers~\cite{Abdullaev2,Abdullaev3}.
On the other hand, for $N_{1}=N_{2}=N/2$ the compacton existence
will depend on $\Omega$ but not on $\sigma$, this is because the $\sigma$
term always involves as a factor a zero amplitude on a site $n_{0}\pm1$
different to the site $n_{0}$. In both cases, however, the SOC terms
will strongly influence the stability, therefore  it is
worth considering in more detail the existence conditions for these
two cases.

\noindent \textbf{Case $\Omega=0.$}

We take $s=0$ in Eq.~\eqref{eq:013} and fix the amplitudes at site
$n_{0}$ as $A_{n_{0}}=a+ib$, $B_{n_{0}}=c+id$, with $a,b,c,d$
reals and by substituting into Eqs.~\eqref{eq:011}, one obtains in
correspondence to sites $n_{0}\pm1$ four equations that are
automatically satisfied if $J_{0}(\alpha(c^{2}+d^{2}))=0$ and $J_{0}(\alpha(a^{2}+b^{2}))=0$,
e.g., if $a,b,c,d$ are taken as
\begin{equation}
a^{2}+b^{2}=N_{1}=\frac{\xi_{m}}{\alpha},\;\;c^{2}+d^{2}=N_{2}=\frac{\xi_{l}}{\alpha},\label{eq:014}
\end{equation}
with $\xi_{m},\xi_{l}$ two different zeros of the Bessel function
$J_{0}$. One can then check that the other two equations, e.g., the
ones for the site $n_{0}$, give the chemical potentials as
\begin{equation}
\mu_{1}=N_{2}\gamma^{(0)}+N_{1}\gamma_{1},\quad\mu_{2}=N_{1}\gamma^{(0)}+N_{2}\gamma_{2}.\label{eq:015}
\end{equation}

\noindent \textbf{Case ${\bf \Omega\neq0.}$}

In this case, we must have $N_{1}=N_{2}=N/2$, so we fix $s=0$ in
Eq.~\eqref{eq:013} and take $A_{n_{0}}=B_{n_{0}}=a+ib$, with $a,b$
reals. One can check that the equations in correspondence to the edges
$n_{0}\pm1$ are automatically satisfied if $J_{0}(\alpha(a^{2}+b^{2}))=0$,
e.g., if $a^{2}+b^{2}=N/2=\xi_{l}/\alpha$, with $\xi_{l}$ a zeros
of $J_{0}$. The other two equations for the site $n_{0}$ can be
solved for the chemical potential only if $\gamma_{1}=\gamma_{2}\equiv\tilde{\gamma}$,
this giving:
\begin{equation}
\mu=\frac{N}{2}(\gamma^{(0)}+\tilde{\gamma})+\Omega.\label{eq:016}
\end{equation}
Note that in this case $\Omega$ enters the $\mu$ expressions simply
as a shift.

\subsection{Two-site SOC-compactons}

For a two-site compacton, $\Omega=0$ will not provide real solutions
for the chemical potentials due to the presence of imaginary terms
proportional to $\sigma$. It is possible however to eliminate such
terms by considering some suitable ansatzes for the amplitudes. For
this, let us fix $s=1$ in Eq.~\eqref{eq:013} and take $A_{n_{0}}=B_{n_{0}+1}=a+ib$,
$B_{n_{0}}=A_{n_{0}+1}=a-ib$, with $a,b$ reals (in-phase solution).
From Eqs.~\eqref{eq:012} one then obtains eight equations in
correspondence of sites $n_{0}-1, n_{0}, n_{0}+1, n_{0}+2$. One
can show that these equations lead to the following solution for the
chemical potential
\begin{equation}
\mu=\left(a^{2}+b^{2}\right)\left(\frac{\Omega-\varGamma}{a^{2}-b^{2}}+\gamma+\gamma^{(0)}\right),\label{eq:017}
\end{equation}
with $a,b$ given by
\begin{equation}
\begin{gathered}a^{2}=\frac{\xi_{l}}{2\alpha}\left(1+\frac{\Omega-\varGamma}{\sqrt{(\varGamma-\Omega)^{2}+\sigma^{2}}}\right),\\
b^{2}=\frac{\xi_{l}}{2\alpha}\left(1-\frac{\Omega-\varGamma}{\sqrt{(\varGamma-\Omega)^{2}+\sigma^{2}}}\right),
\end{gathered}
\label{eq:018}
\end{equation}
and $\xi_{l}$ denoting a zero of $J_{0}$. Using Eqs.~\eqref{eq:018}
one can rewrite Eqs.~\eqref{eq:017} in a more explicit form
\begin{equation}
\mu=\frac{N}{4}\left(\gamma+\gamma^{(0)}\right)+\sqrt{\sigma^{2}+(\varGamma-\Omega)^{2}},\label{eq:019}
\end{equation}
showing the full dependence on parameters $\varGamma,\sigma,\Omega$
and on the total numbers of atoms $N$.

Similarly, one can obtain out-of-phase two-site compactons solutions
by assuming the ansatz $A_{n_{0}}=-B_{n_{0}+1}=a+ib$,  $B_{n_{0}}=-A_{n_{0}+1}=a-ib$.
In this case, the expressions for $\mu,a,b$ are the same as in Eqs.~\eqref{eq:017}~-~\eqref{eq:019}, except for the replacement $\varGamma\rightarrow-\varGamma$
in all equations and for the change of sign in front of the fractions
appearing inside the parenthesis in Eqs.~\eqref{eq:018}.

\begin{figure*}
\includegraphics[scale=0.8]{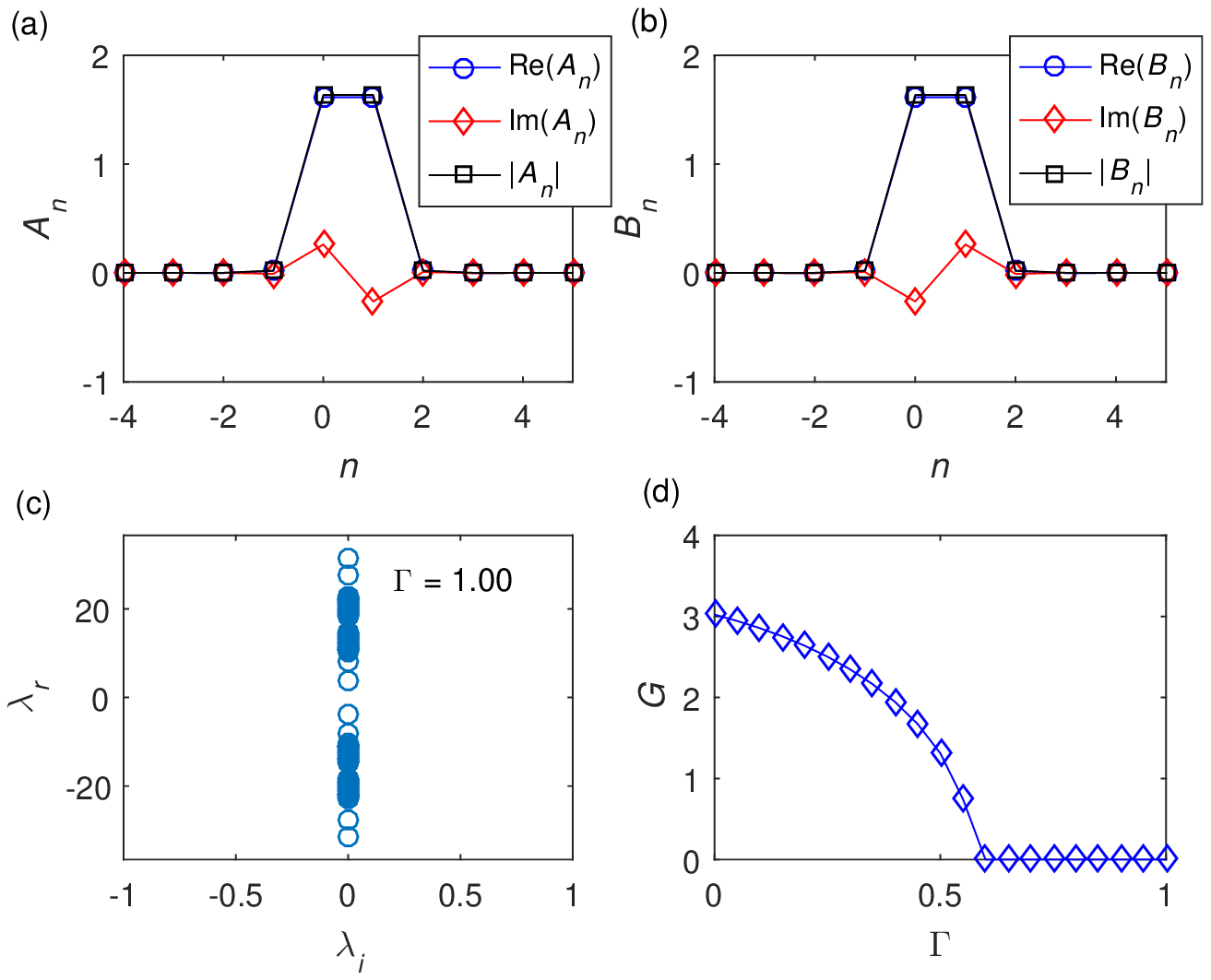}

\caption{Two-site in-phase compacton for case $\gamma_{1}=\gamma_{2}=1$, $\gamma^{(0)}=4$,
$\gamma^{(1)}=1$, $\omega=1$, $\Omega=4$ and $\sigma=1$}.

\label{Figure05}
\end{figure*}

\begin{figure*}[p]
\includegraphics[scale=0.8]{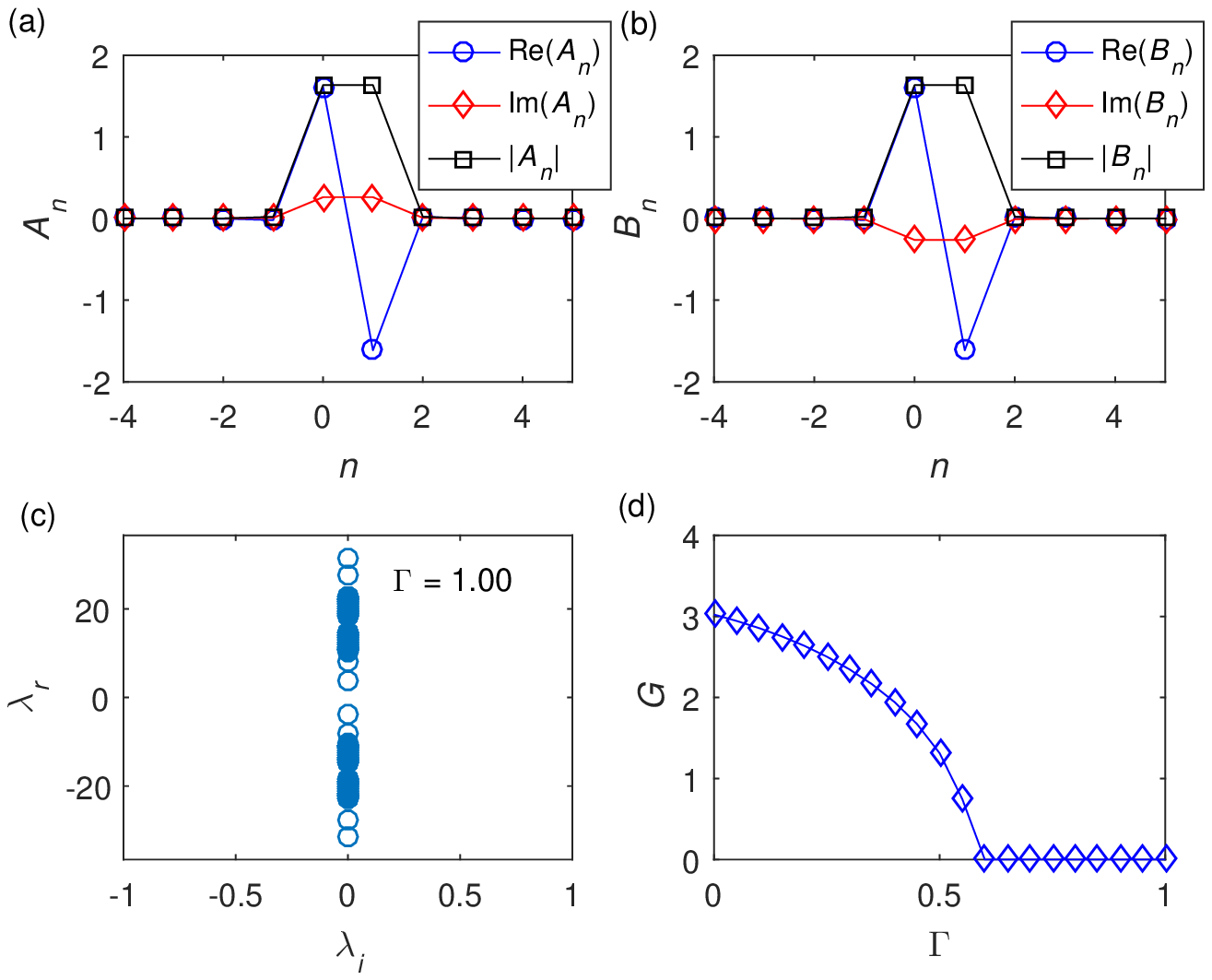}

\caption{Two-site out-phase compacton for case $\gamma_{1}=\gamma_{2}=-1$,
$\gamma^{(0)}=-4$,$\gamma^{(1)}=1$, $\omega=1$, $\Omega=-4$ and
$\sigma=1$}

\label{Figure06}
\end{figure*}

\begin{figure*}[p]
\includegraphics[scale=0.8]{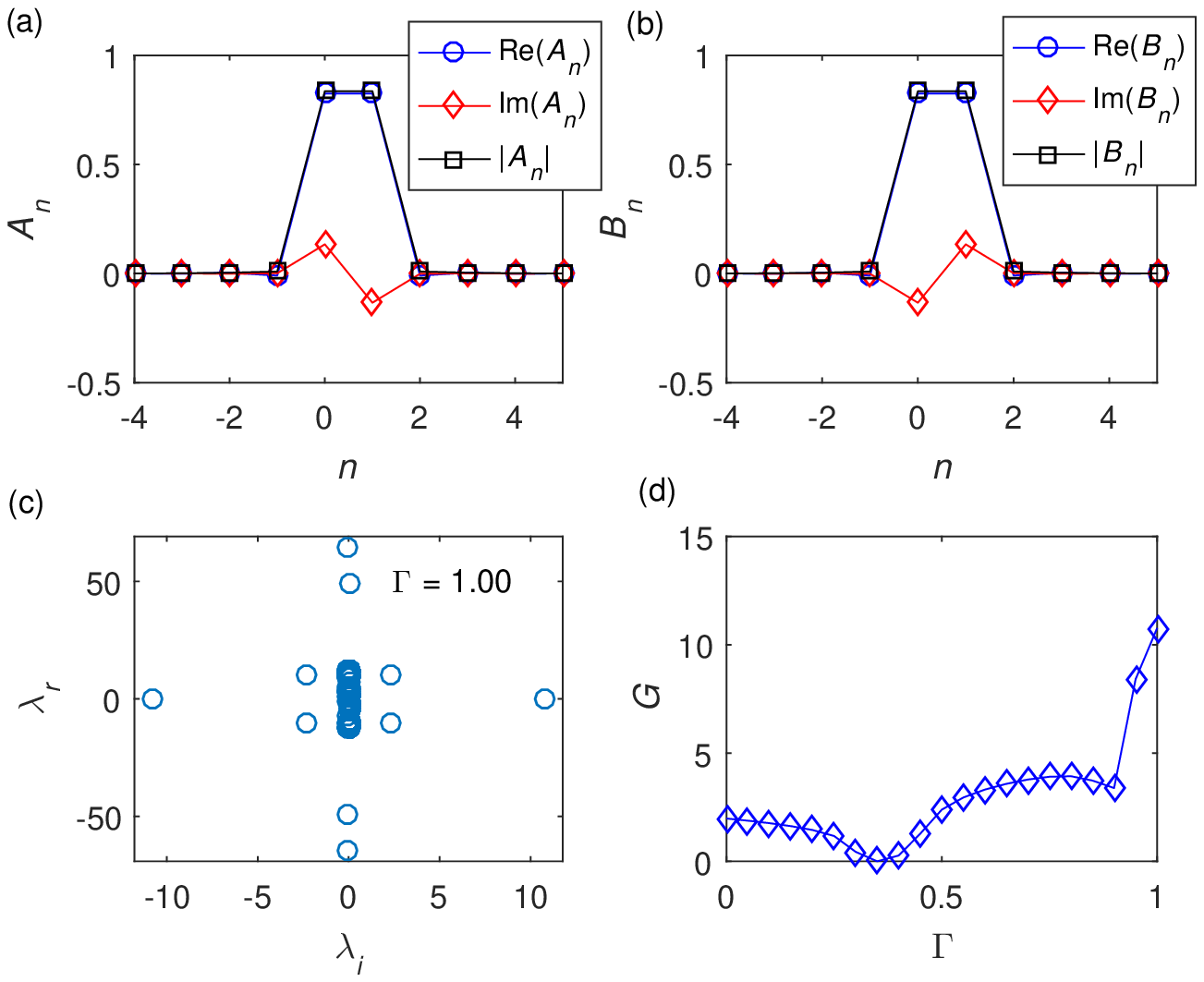}

\caption{Two-site in-phase compacton for case $\gamma_{1}=\gamma_{2}=1$, $\gamma^{(0)}=4$,
$\gamma^{(1)}=8.0$, $\omega=1$, $\Omega=4$ and $\sigma=1$}
\label{Figure07}
\end{figure*}

\begin{figure*}
\includegraphics[scale=0.9]{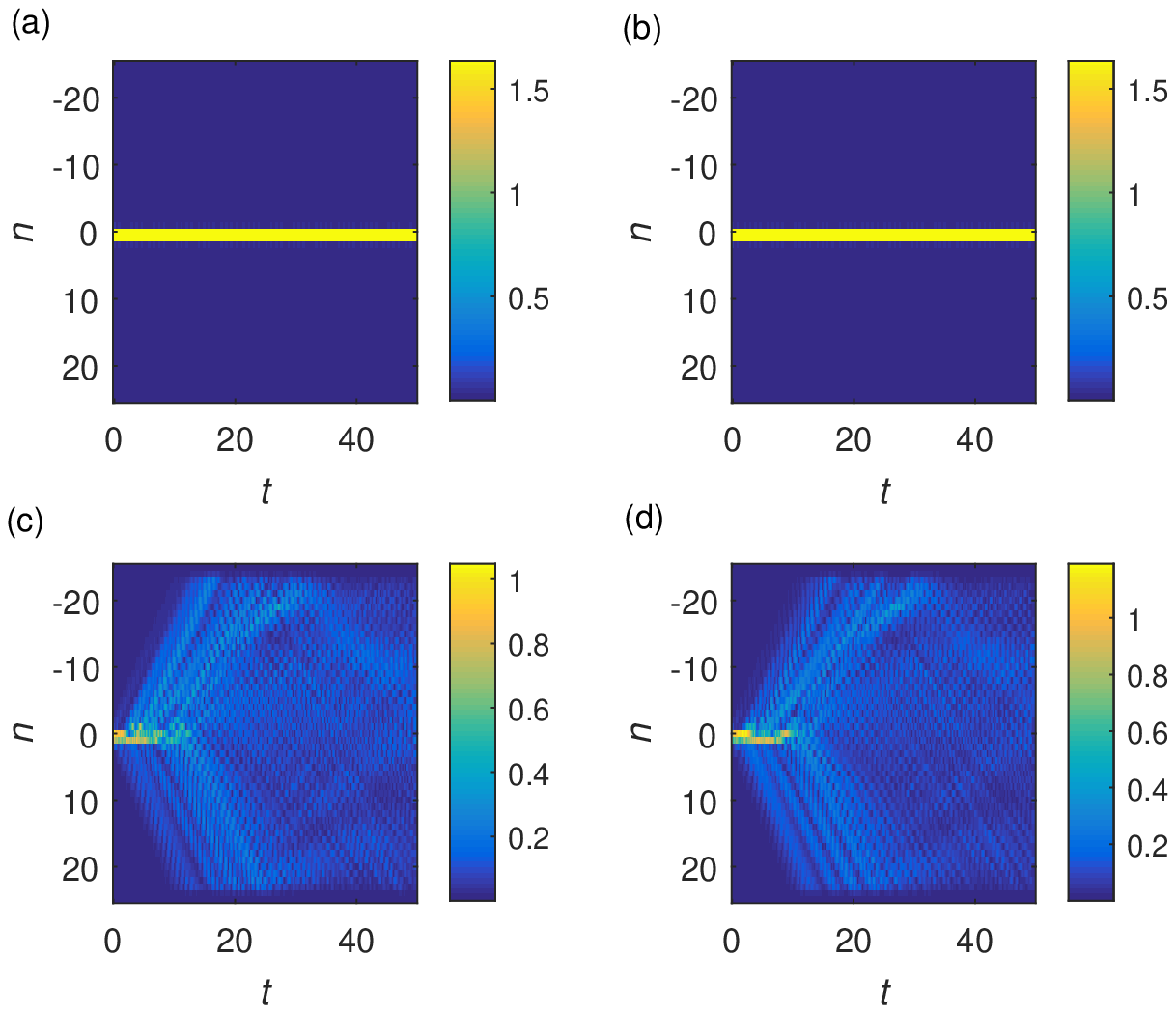}

\caption{Space-time evolution of the two-site compactons obtained from numerical integrations of  Eq.~\eqref{eq:004}
with $\epsilon=0.01$:  (a) $\left|u_{n}\right|$ and (b) $\left|v_{n}\right|$
for the stable case at $\Gamma=1$  in Fig.\ref{Figure05}; (c) $\left|u_{n}\right|$
and (d) $\left|v_{n}\right|$ for unstable case at  $\Gamma=1$  in Fig.\ref{Figure07}.}
\label{Figure08}
\end{figure*}

\subsection{Multi-site SOC-compactons}

In principle, the above analysis, being completely general, can be
applied to SOC-compactons of arbitrary size. The equations to solve,
however, in these cases become much more involved for an exact analysis
and one must recourse to numerical methods.
An example of this
is given for a three-site compacton in the next section.

\section{Numerical results}

In order to check the above results, we will use both linear stability
analysis and direct numerical integrations of the
original discrete CGPE equations in Eq.~\eqref{eq:004}. For the
linear stability analysis, the standard procedure is employed by considering
the ansatz with perturbation of the form:
\begin{equation}
\begin{gathered}U_{n}=[A_{n}+\varepsilon(a_{n}e^{-i\lambda t}+b_{n}e^{i\lambda^{*}t})]e^{-i\mu t},\\
V_{n}=[B_{n}+\varepsilon(c_{n}e^{-i\lambda t}+d_{n}e^{i\lambda^{*}t})]e^{-i\mu t},
\end{gathered}
\label{eq:020}
\end{equation}
where $\lambda=\lambda_{r}+i\lambda_{i}$ denotes the linearization
eigenvalue and $\varepsilon\ll{1}$. Substituting Eq.~\eqref{eq:020}
in Eqs.~\eqref{eq:023}, \eqref{eq:024} and taking only the
terms with $O\left(\varepsilon\right)$ (linear terms of $\varepsilon$),
one obtains an eigenvalue problem for $\lambda$ that can be solved numerically. Ideally, the perturbation part in Eq.~\eqref{eq:020} will
remain small (stable) if all imaginary eigenvalues $\lambda_{i}$
are zeros or negligibly small, for gain $G=\max\left(\left|\lambda_{i}\right|\right)\simeq0$.

Typically, the  results of the linear stability analysis of SOC-compactons are further checked by direct integrations of both the averaged and original equations Eq.~\eqref{eq:004}. Since they are found in agreement, however, results in the following will be reported only for the original equations.

\subsection{One-site SOC-compactons}

As discussed above, the existence condition for one-site SOC-compactons, apart
possible shifts of the chemical potential by $\Omega$, are the same as in the
absence of SOC. This, however, does not imply
that there is no effects of the SOC on compactons  because
both $\sigma$ and $\Omega$ can strongly influence their stability.

We show this by comparing results of the linear stability analysis
in the case of absence of SOC , i.e. $\sigma=0,\Omega=0$, (see Fig.~\ref{Figure01})
with the ones obtained in the presence of SOC (see Figs.\ref{Figure02}
and~\ref{Figure03}). From Fig.~\ref{Figure01} we see that the linear
spectrum has a gain $G$ that is practically zero for all $\varGamma\in\left[0,1\right]$,
meaning that the solutions are stable, a fact that is already known
from previous results~\cite{Abdullaev3}. By increasing $\sigma$
away from zero but retaining $\Omega=0$, we find
that the solution becomes unstable (see panels a, b
of Fig.~\ref{Figure02}) for the case $\sigma=1,\Omega=0$. This instability,
however, is physically not very significative because the SOC implementation
usually requires both $\sigma,\Omega$ to be nonzero. By increasing
also $|\Omega|$ away from zero, the stability can be restored almost
for the whole range in $\varGamma$ of the zero SOC case, as one can
see from panels (c, d) of Fig.~\ref{Figure02}, for the case $\sigma=1,\Omega=-2$.
Similar results are obtained for different choices of the system
parameters and for different compacton amplitudes. In general, we
have that stable one-site SOC-compactons can exist in wide regions
of the parameter space.

We have also investigated the effects of the nonlinearity management parameters
$\gamma^{(0)}$, $\gamma^{(1)}$, on the stability of one-site
SOC-compactons. Note from Eq.~\eqref{eq:014} that the compacton amplitude
is inversely proportional to the value of $\gamma^{(1)}$, despite
$\gamma^{(1)}$ alone does not affect the stability. Remarkably,
the solutions become significantly stable when $\left|\gamma^{(0)}\right|$
is much larger than $\gamma^{(1)}$ as demonstrated in Fig.~\ref{Figure03}(a,
b) and, oppositely, the solutions become unstable when $\left|\gamma^{(0)}\right|$
is much lower than $\gamma^{(1)}$, as portrayed in Fig.~\ref{Figure03}
(c, d). The above linear stability results are further confirmed by
nonlinear stability analysis achieved via numerical time-integrations
of the nonlinear system in Eq.~\eqref{eq:004}, as one can see from
Fig.~\ref{Figure04}.

\subsection{Two-site SOC-compactons}

For multi-site compactons, we must necessarily to have the condition $\Omega\neq0$, i.e. equal chemical potentials for the two BEC components and  this implies  equal intra-species interactions. Let us  consider first the case of an in-phase SOC-compacton with amplitudes
$A_{0} = B_{1}=a+ib, A_{1}= B_{0}=a-ib,$ with $a=\sqrt{\xi_{1}/2\alpha}$,
$b=\sqrt{\xi_{1}/\alpha-a^{2}}$. We fix parameters as:
$\gamma_{1}=\gamma_{2}=1$, $\gamma^{(0)}=4$, $\gamma^{(1)}=1$,
$\omega=1$, $\Omega=4$, $\sigma=1$. Note that all the   nonlinear
parameters here are positive, i.e  interactions are repulsive. In panels a,b of  Fig.~\ref{Figure05}   we show  the real and imaginary parts versus $n$ of the first and second component $A_n,  B_n$ ,  respectively, of a  two-site in-phase compactons, while  in the bottom panels (c,d) their stability analysis is reported as a function of the parameter $\Gamma$. From panel d) it is clear that for $\Gamma$ less than  $\approx 0.6$  the gain $G$ is positive and  the solution is unstable,  while for all $\Gamma \geq 0.6$ the gain reduces to zero and the solution becomes stable. The  panel c of the figure  just shows the stability in terms of the linear eigenvalues  at  $\Gamma=1$ from which we see that they are all real numbers.

Similarly, for an out-of-phase two-site compacton we fix  the amplitudes as:
$A_{0}=-B_{1}=a+ib$, $B_{0}=-A_{1}=a-ib$, with  the same  $a$, $b$  of the  previous case.  Parameters are fixed  as $\gamma_{1}=\gamma_{2}=-1$, $\gamma^{(0)}=-4$, $\gamma^{(1)}=1$,
$\omega=1$, $\Omega=-4$, $\sigma=1$. Results,  are reported in Fig.~\ref{Figure06},  where panels (a, b) depict the solutions of typical out-of-phase profiles and panels   (c,d) show the results from the stability analysis. Notice that  despite the type of interactions is changed from repulsive to attractive, the range of stability is the same as for the  in-phase case. Thus, independently on the sign of the interaction,  in the interval $0.6\leq\varGamma\leq1$ both the in-phase and the out-of phase compactons  appear to be stable. This  is quite interesting because it shows SOC-compacton bistability in contrast with usual inter-site breathers (i.e. localized out-of-phase solutions with tails) which  are known to be  unstable in conventional (i.e. in absence of  SNM)   SOC-BEC mixtures~\cite{Belicev}.

The stability of the solution also depends on the ratio $\eta=\gamma^{(0)}<\gamma^{(1)}$ between the dc and ac strengths of the inter-site interaction, with the two-site SOC-compacton  becoming unstable for $\eta < 1$. This is shown in Fig.\ref{Figure07} for the case of the same  two-site in-phase compacton depicted  in Fig. \ref{Figure05} but with $\eta=0.5$.

Finally, we find the results of the linear stability analysis to be in very good agreement with those obtained from  direct numerical integrations of Eq.~\eqref{eq:004}. An example of this is shown in Fig.\ref{Figure08}  where the  stable and unstable time evolution of the two-site in-phase compactons of  Fig.s \ref{Figure05} and  \ref{Figure07} at $\Gamma=1$, are reported.

\begin{figure*}
\includegraphics[scale=0.8]{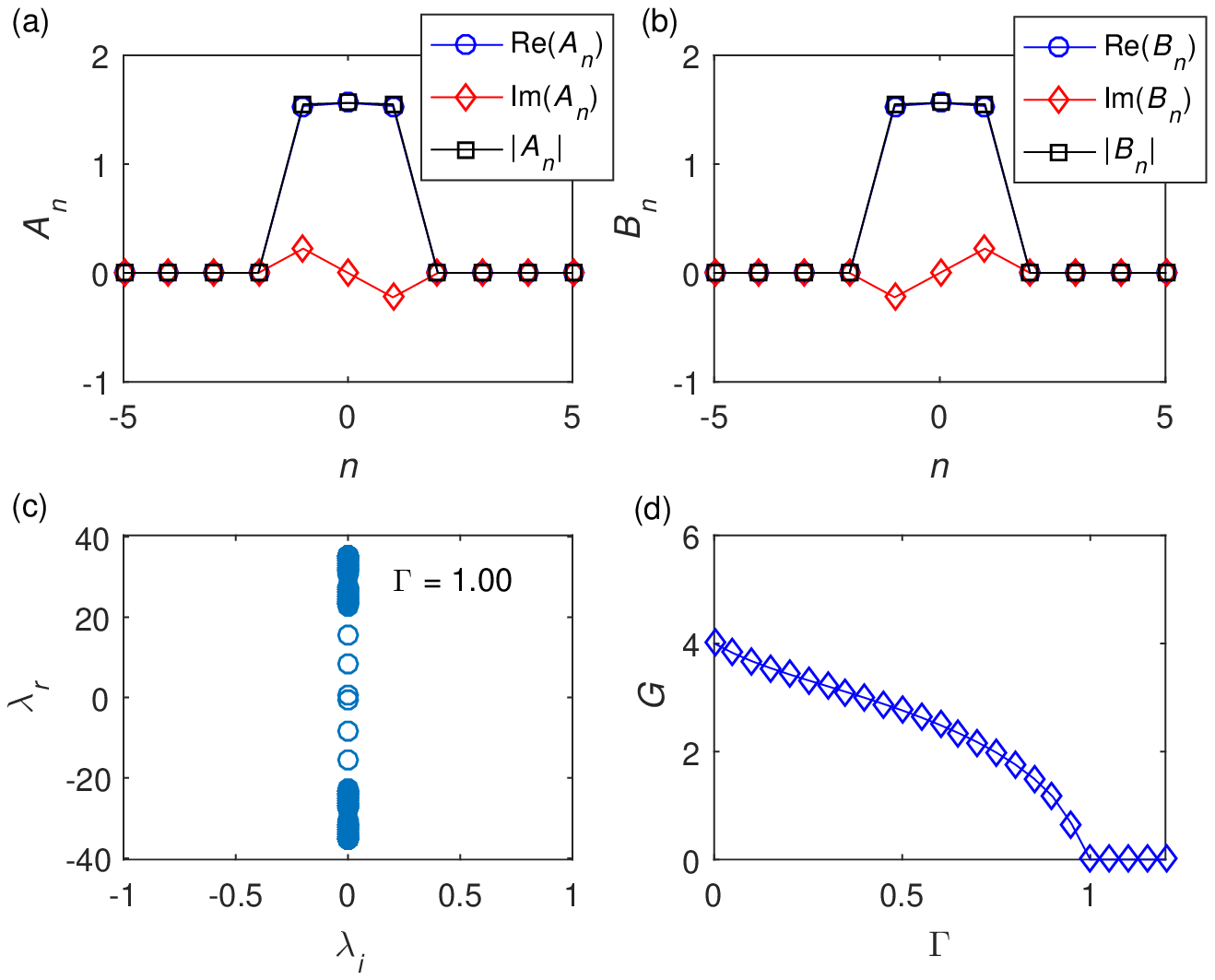}

\caption{Three-site compacton for case $\gamma_{1}=\gamma_{2}=1$, $\gamma^{(0)}=10$,
$\gamma^{(1)}=1$, $\omega=1$, $\Omega=4$ and $\sigma=1$.}

\label{Figure09}
\end{figure*}

\begin{figure*}
\includegraphics[scale=0.8]{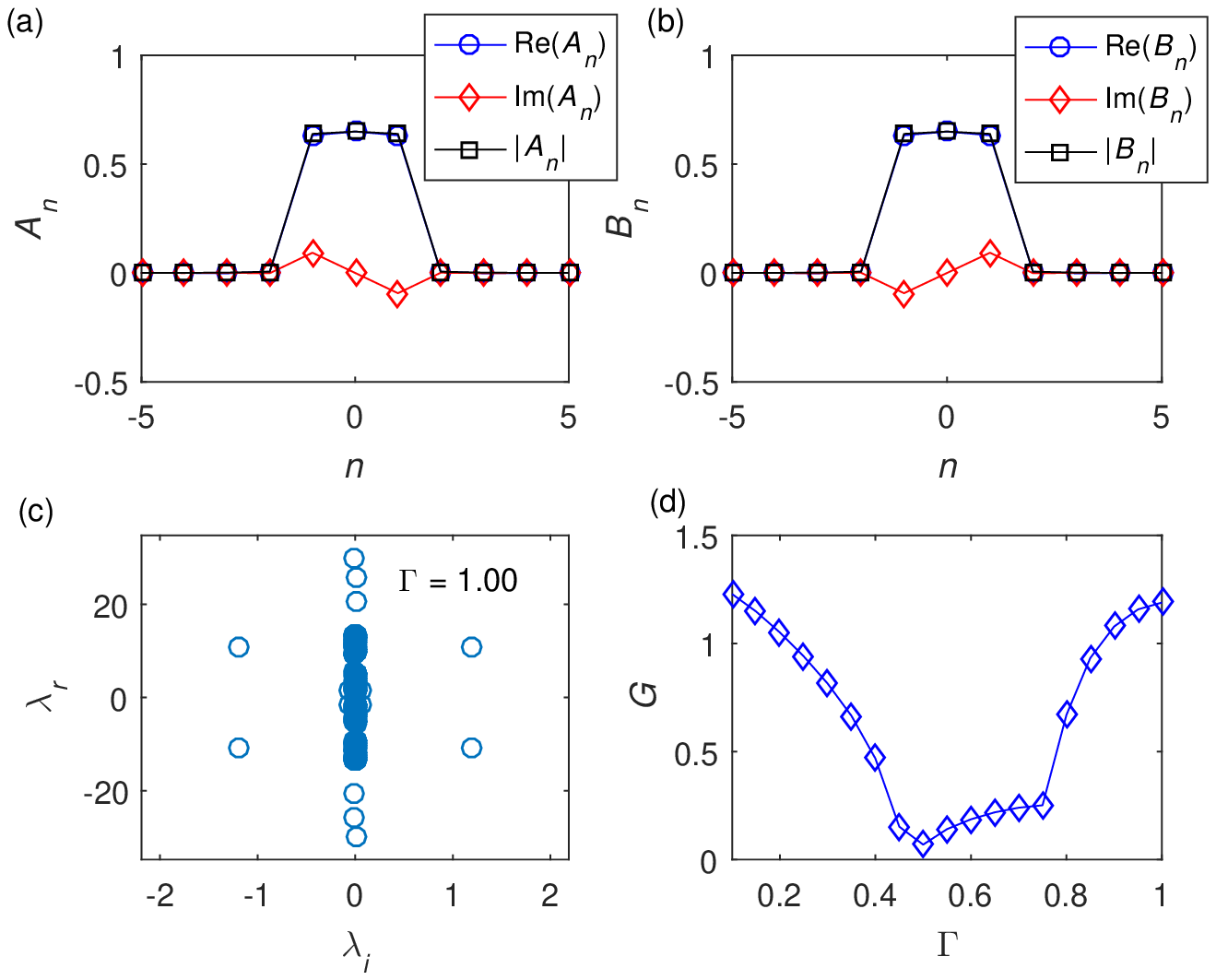}

\caption{Three-site compacton for case $\gamma_{1}=\gamma_{2}=1$, $\gamma^{(0)}=10$,
$\gamma^{(1)}=6.0$, $\omega=1$, $\Omega=4$ and $\sigma=1$.}

\label{Figure10}
\end{figure*}

\begin{figure*}
\includegraphics[scale=0.8]{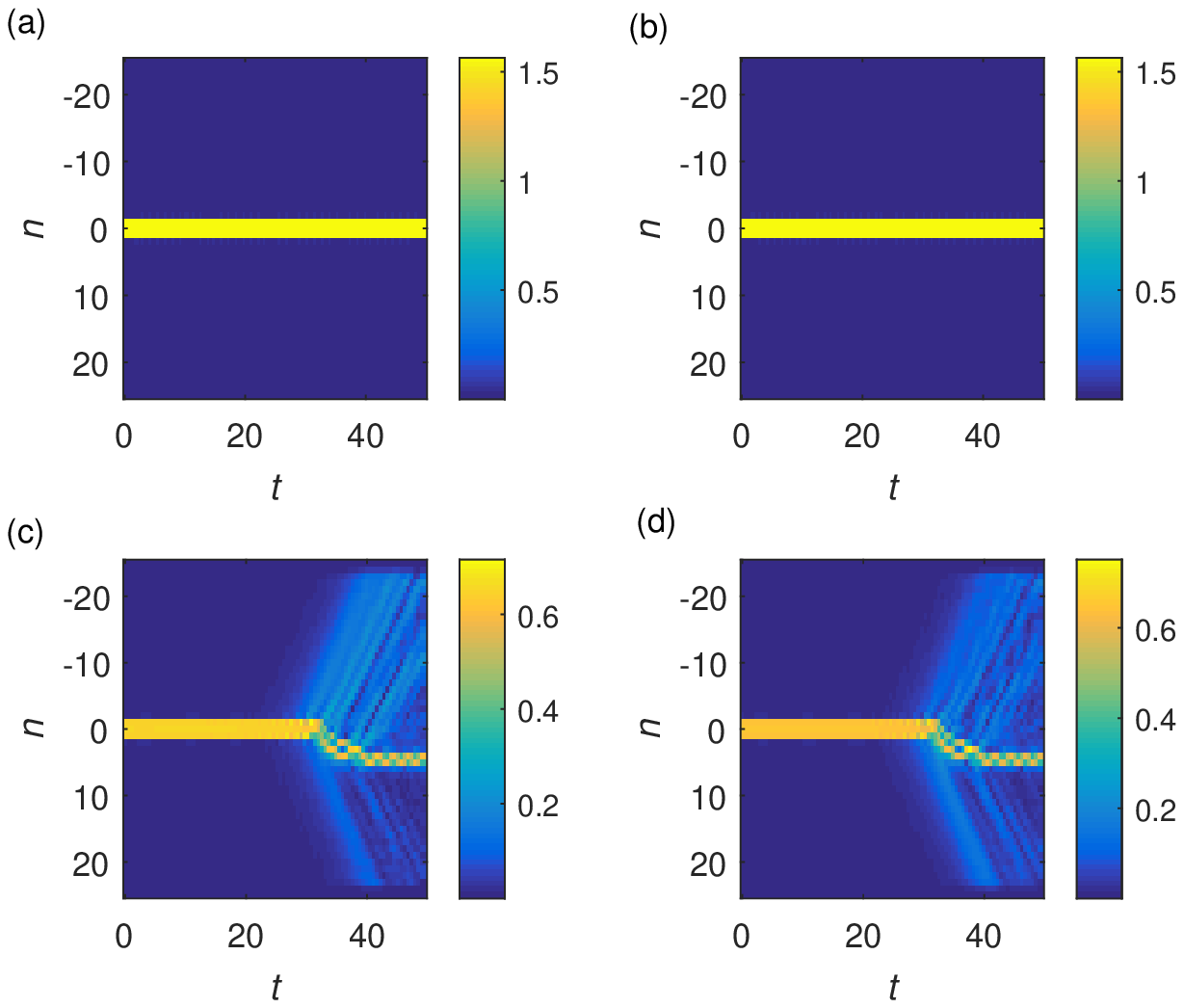}

\caption{Space-time evolution of the three-site compactons obtained from numerical integrations of Eq.~\eqref{eq:004}
with $\epsilon=0.01$,  (a) $\left|u_{n}\right|$ and (b) $\left|v_{n}\right|$
for the stable case at $\Gamma=1$  in Fig.\ref{Figure09}; (c) $\left|u_{n}\right|$
and (d) $\left|v_{n}\right|$ for unstable case at $\Gamma=1$  in Fig.\ref{Figure10}.}

\label{Figure11}
\end{figure*}

\subsection{Three-site SOC-compactons}
Three-site SOC-compacton solutions are obtained from Eq.~\eqref{eq:013} with $s=2$  by taking
$\mu_{1}=\mu_{2}=\mu$ and assuming the following  ansatz for the solution: $A_{n_{0}}=a+ib$, $B_{n_{0}}=a-ib$, $A_{n_{0}\pm1}=c+id$ and $B_{n_{0}\pm1}=c-id$, with  $a, b, c,d$ real numbers.  Direct substitution of the ansatz  into Eq.~\eqref{eq:012} provides  ten equations which are  reduced to six (for instance only those for sites $n_{0}-2,n_{0}-1,n_{0}$)  thanks to the symmetry of the solution  around the central site $n_0$. From the equations for the site $n_0-2$ one obtains the zero tunneling conditions  for $A_{n_{0}\pm 2}, B_{n_{0}\pm 2}$ at the edges of the compacton in terms of  roots of the Bessel
function as in Eq.~\eqref{eq:014}. Substituting these conditions into  the equation relative to  sites $ n_0- 1$  one obtains two equations that can be combined into the  following expression for the   chemical potential:
\begin{eqnarray*}
\mu & = & -\frac{1}{Re\left(w_{2}\right)}\left[\alpha\varGamma\left(Re\left(w_{2}\bar{z}_{1}w_{1}\right)\right.\right.\\
 &  & \left.+Re\left(w_{2}z_{1}\bar{w}_{1}\right)\right)J_{1}\left(\alpha\left(\left|z_{2}\right|^{2}-\left|w_{2}\right|^{2}\right)\right)\\
 &  & +\varGamma Re\left(z_{2}\right)J_{0}\left(\alpha\left(\left|z_{1}\right|^{2}-\left|w_{1}\right|^{2}\right)\right)\\
 &  & -\Omega Re\left(w_{1}\right)-\gamma\left|w_{1}\right|^{2}Re\left(w_{2}\right)\\
 &  & -\gamma^{\left(0\right)}\left|w_{1}\right|^{2}Re\left(w_{2}\right)\\
 &  & -\alpha\sigma J_{1}\left(\alpha\left(\left|z_{2}\right|^{2}-\left|w_{2}\right|^{2}\right)\right)\cdot\\
 &  & \left(Im\left(w_{2}\bar{z}_{1}w_{1}\right)-Im\left(w_{2}z_{1}\bar{w}_{1}\right)\right)\\
 &  & -\sigma J_{0}\left(\alpha\left(\left|z_{1}\right|^{2}-\left|w_{1}\right|^{2}\right)\right)\left.Im\left(z_{2}\right)\right].
\end{eqnarray*}
Here we used $\gamma_{1}=\gamma_{2}=\gamma$ and denoted $z_{1}=a+i b, z_{2} = a - i b, w_{1}=c + i d, w_{2}=c-id$. Lastly, from the equations for site $n_{0}$ one gets:
\begin{eqnarray*}
\mu z_{j+1}+2\alpha\varGamma z_{j+1}\left(\bar{w}_{2-j}z_{2-j}+w_{2-j}\bar{z}_{2-j}\right)\cdot\\
J_{1}\left(\alpha\left(\left|w_{j+1}\right|^{2}-\left|z_{j+1}\right|^{2}\right)\right)\\
+\alpha\varOmega\left(z_{2-j}\left|z_{j+1}\right|^{2}+z_{2-j}z_{j+1}^{2}\right)\cdot\\
J_{1}\left(\alpha\left(\left|z_{j+1}\right|^{2}-\left|z_{2-j}\right|^{2}\right)\right)\\
+2\varGamma w_{j+1}J_{0}\left(\alpha\left(\left|w_{2-j}\right|^{2}-\left|z_{2-j}\right|^{2}\right)\right)\\
-\varOmega z_{2-j}J_{0}\left(\alpha\left(\left|z_{j+1}\right|^{2}-\left|z_{2-j}\right|^{2}\right)\right)\\
-\gamma\left|z_{j+1}\right|^{2}z_{j+1}-\gamma^{\left(0\right)}\left|z_{2-j}\right|^{2}z_{j+1} & = & 0,
\end{eqnarray*}
with  $j=0,1$. These nonlinear equations together with the previous ones should be solved numerically   for $a,b,c,d$ giving the the compacton amplitudes of the two components at sites $n_0, n_0 \pm 1$.

Stability patterns for three-site compactons are in general similar to the ones previously investigated but  for  $\sigma=1$ or larger it is more difficult to obtain stable solutions unless $\varGamma\geq1$, as shown in  Fig.~\ref{Figure09}. Another distinctive feature  in this case is the absence of stable  solutions  $\gamma^{(0)}<\gamma^{(1)}$, and even for $\gamma^{(0)}>\gamma^{(1)}$,  if the difference is not sufficiently large, the solution might become unstable as depicted in Fig.~\ref{Figure10}.
Lastly,  in the top and bottom panels of  Fig.~\ref{Figure11} we report the time evolution obtained from numerical integrations of Eq.~\eqref{eq:004} by  initial conditions the solutions at $\Gamma=1$ depicted in  Fig.~\ref{Figure09} and Fig.~\ref{Figure10}, for which the linear stability analysis predicts stability and instability, respectively. As one can see from panels (a,b) and (c,d) of
Fig.~\ref{Figure11} these predictions are fully confirmed.

\section{Conclusions}

In this paper we have investigated the existence and stability properties
of compactons matter
waves excitations of binary BEC mixtures with spin-orbit coupling
trapped in an OL, subjected to SNM. We considered for simplicity an inter-SOC system (i.e.
with an off-diagonal Rabi term) and assumed both the OL to be deep
enough to justify the tight-binding approximation and the strength
and the frequency of the modulation to be large enough to use the averaging
method.

Within this model, we derived effective averaged
equations of motion for the matter wave complex amplitudes in the
form of a two coupled discrete NLS and
demonstrated that the SNM introduces a local density imbalance dependence  not only
in the tunneling parameter $\Gamma$ but also in the SOC parameters $\sigma$ and
$\Omega$.
This density dependence of the SOC  parameters persists in the system
independently on the SOC-compacton existence.

The dependence of  both $\sigma$ and $\Omega$ on the local density imbalance, however,  give rise
to conditions for stable SOC-compactons existence that are more restrictive than
the ones obtained in the absence of SOC.
In general, the stationarity condition requires the equality of the chemical potentials of the
two components, a fact that can be satisfied when the number of atoms
of the two species are equal as well as their intra-species nonlinearities.
We have also investigated the stability of SOC-compactons versus the
system parameters, i.e. the SOC strength $\sigma$, the Rabi frequency
$\Omega$, the coupling or hopping constant $\varGamma$ and the ratio
between the nonlinear coefficients $\gamma^{(0)},\gamma^{(1)}$. This
has been done both using a linear stability analysis of the
averaged equations and by direct numerical time integrations of the
original nonlinear PDE equations. Numerical simulations show that
by keeping the other parameters fixed, an enlargement of the stability
region of SOC-compactons can be achieved by decreasing $\sigma$ or
by increasing $\Omega$. Stability seems also more easily obtainable
for $\left|\gamma^{(0)}\right|>\left|\gamma^{(1)}\right|$ and large
values of $\varGamma$.

From this, we conclude that while  on one side the SOC restricts
the parameter range for stationary SOC-compacton existence, on the
other side it gives a more stringent signature of their occurrence.
Indeed, they should appear when the intra-species interactions and
the number of atoms in the two components are perfectly balanced (or
close to being balanced for metastable cases). This suggests SOC-compactons
as possible tools for indirect measurements of the number of atoms,
and/or the intra-species interactions, in experiments.

\section*{ACKNOWLEDGMENTS}

M.S.A. H. and L. A. T. acknowledge support from Ministry of Higher
Education, Malaysia under research grant no. FRGS 16-013-0512. MS
wishes to acknowledge the International Islamic University of Malaysia
in Kuantan for partial support and for the hospitality received during
the period this work was started.

\section{APPENDIX: Averaged equations and Hamiltonian structure}

Below we provide the derivation of the averaged system Eq.~(\ref{eq:010}).
By substituting Eq.
(\ref{eq:007}) into Eq.
(\ref{eq:004}) we get:
\begin{align}
iU_{n,t}= & -i\varGamma U_{n}(\langle\Lambda\bar{X}_{1}^{+}\rangle V_{n}-\langle\Lambda X_{1}^{+}\rangle V_{n}^{*})\nonumber \\
 & -\sigma U_{n}(\langle\Lambda\bar{X}_{1}^{-}\rangle V_{n}+\langle\Lambda X_{1}^{-}\rangle V_{n}^{*})\nonumber \\
 & +i\Omega(\langle\Lambda\bar{X}_{0}^{+}\rangle V_{n}\vert U_{n}\vert^{2}-\langle\Lambda X_{0}^{-}\rangle V_{n}^{*}U_{n}^{2})\nonumber \\
 & -\varGamma\langle X_{2}^{+}\rangle+i\sigma\langle X_{2}^{-}\rangle+\Omega\langle X_{0}^{-}\rangle V_{n}\nonumber \\
 & +\gamma_{1}\vert U_{n}\vert^{2}U_{n}+\gamma^{(0)}\vert V_{n}\vert^{2}U_{n},\label{eq:021}\\
iV_{n,t}= & -i\varGamma V_{n}(\langle\Lambda\bar{X}_{2}^{+}\rangle U_{n}-\langle\Lambda X_{2}^{+}\rangle U_{n}^{*})\nonumber \\
 & +\sigma V_{n}(\langle\Lambda\bar{X}_{2}^{-}\rangle U_{n}+\langle\Lambda X_{2}^{-}\rangle U_{n}^{*})\nonumber \\
 & +i\Omega(\langle\Lambda\bar{X}_{0}^{+}\rangle U_{n}\vert V_{n}\vert^{2}-\langle\Lambda X_{0}^{-}\rangle U_{n}^{*}V_{n}^{2})\nonumber \\
 & -\varGamma\langle X_{1}^{+}\rangle-i\sigma\langle X_{1}^{-}\rangle+\Omega\langle X_{0}^{+}\rangle U_{n}\nonumber \\
 & +\gamma_{2}\vert V_{n}\vert^{2}V_{n}+\gamma^{(0)}\vert U_{n}\vert^{2}V_{n}.\label{eq:022}
\end{align}
where
\begin{align*}
X_{2}^{\pm} & =U_{n+1}e^{i\Lambda(\vert{V_{n+1}}\vert^{2}-\vert{V_{n}}\vert^{2})}\pm U_{n-1}e^{i\Lambda(\vert{V_{n-1}}\vert^{2}-\vert{V_{n}}\vert^{2})},\\
X_{1}^{\pm} & =V_{n+1}e^{i\Lambda(\vert{U_{n+1}}\vert^{2}-\vert{U_{n}}\vert^{2})}\pm V_{n-1}e^{i\Lambda(\vert{U_{n-1}}\vert^{2}-\vert{U_{n}}\vert^{2})},\\
X_{0}^{\pm} & =e^{\pm i\Lambda(\vert{U_{n}}\vert^{2}-\vert{V_{n}}\vert^{2})}.
\end{align*}
and $\langle\cdot\rangle$ is the rapid modulated term to be averaged.
The averaged terms can be evaluated as:
\begin{gather*}
\langle\Lambda X_{1}^{\pm}\rangle=V_{n+1}\langle\Lambda e^{i\Lambda\theta_{1}^{+}}\rangle\pm V_{n-1}\langle\Lambda e^{i\Lambda\theta_{1}^{-}}\rangle,\\
\langle\Lambda X_{2}^{\pm}\rangle=U_{n+1}\langle\Lambda e^{i\Lambda\theta_{2}^{+}}\rangle\pm U_{n-1}\langle\Lambda e^{i\Lambda\theta_{2}^{-}}\rangle,\\
\langle X_{1}^{\pm}\rangle=V_{n+1}\langle e^{i\Lambda\theta_{1}^{+}}\rangle\pm V_{n-1}\langle e^{i\Lambda\theta_{1}^{-}}\rangle,\\
\langle X_{2}^{\pm}\rangle=U_{n+1}\langle e^{i\Lambda\theta_{2}^{+}}\rangle\pm U_{n-1}\langle e^{i\Lambda\theta_{2}^{-}}\rangle,\\
\langle\Lambda X_{0}^{\pm}\rangle=\langle\Lambda e^{\pm{i}\Lambda\theta_{0}}\rangle,\quad\langle X_{0}^{\pm}\rangle=\langle e^{\pm{i}\Lambda\theta_{0}}\rangle,
\end{gather*}
with $\theta_{1}^{\pm}=\vert U_{n\pm1}\vert^{2}-\vert U_{n}\vert^{2}$,
$\theta_{2}^{\pm}=\vert V_{n\pm1}\vert^{2}-\vert V_{n}\vert^{2}$
and $\theta_{0}=\vert U_{n}\vert^{2}-\vert V_{n}\vert^{2}$ and furthermore
\begin{align*}
\langle\Lambda e^{\pm{i}\Lambda\theta_{j}^{\pm}}\rangle & =\frac{\omega}{2\pi}\int_{0}^{2\pi/\omega}\frac{\gamma^{(1)}}{\omega}\sin(\omega\tau)e^{\pm{i}\frac{\gamma^{(1)}}{\omega}\theta_{j}^{\pm}\sin(\omega\tau)}\ d\tau\\
 & =\pm{i}\alpha J_{1}(\alpha\theta_{j}^{\pm}),\;\;\;\;\;\langle e^{\pm{i}\Lambda\theta_{j}^{\pm}}\rangle=J_{0}(\alpha\theta_{j}^{\pm}),\\
\langle\Lambda e^{\pm{i}\Lambda\theta_{0}}\rangle & =\pm i\alpha{J_{1}}(\alpha\theta_{0}),\;\;\;\;\;\langle e^{\pm{i}\Lambda\theta_{0}}\rangle=J_{1}(\alpha\theta_{0}),
\end{align*}
where $\alpha=\gamma^{(1)}/\omega$ and $J_{0}$, $J_{1}$ are the
Bessel functions of order 0 and 1 respectively. Substituting the averages
into Eqs.\eqref{eq:021} and \eqref{eq:022} yields:
\begin{align}
i & U_{n,t}=-\varGamma\Big\{\alpha U_{n}\Big[(V_{n+1}^{*}V_{n}+V_{n+1}V_{n}^{*})J_{1}(\alpha\theta_{1}^{+})+\nonumber \\
 & (V_{n-1}^{*}V_{n}+V_{n-1}V_{n}^{*})J_{1}(\alpha\theta_{1}^{-})\Big]+U_{n+1}J_{0}(\alpha\theta_{2}^{+})+\nonumber \\
 & U_{n-1}J_{0}(\alpha\theta_{2}^{-})\Big\}+\Big\{\gamma_{1}\vert U_{n}\vert^{2}+\gamma^{(0)}\vert V_{n}\vert^{2}\Big\} U_{n}+\label{eq:023}\\
 & i\sigma\Big\{\alpha U_{n}\Big[(V_{n+1}^{*}V_{n}-V_{n+1}V_{n}^{*})J_{1}(\alpha\theta_{1}^{+})-(V_{n-1}^{*}V_{n}-\nonumber \\
 & V_{n-1}V_{n}^{*})J_{1}(\alpha\theta_{1}^{-})\Big]+U_{n+1}J_{0}(\alpha\theta_{2}^{+})-U_{n-1}J_{0}(\alpha\theta_{2}^{-})\Big\}\nonumber \\
 & -\Omega\Big\{\alpha J_{1}(\alpha\theta_{0})(V_{n}\vert U_{n}\vert^{2}+V_{n}^{*}U_{n}^{2})-V_{n}J_{0}(\alpha\theta_{0})\Big\},\nonumber
\end{align}
\begin{align}
i & V_{n,t}=-\varGamma\Big\{\alpha V_{n}\Big[(U_{n+1}^{*}U_{n}+U_{n+1}U_{n}^{*})J_{1}(\alpha\theta_{2}^{+})+\nonumber \\
 & (U_{n-1}^{*}U_{n}+U_{n-1}U_{n}^{*})J_{1}(\alpha\theta_{2}^{-})\Big]+V_{n+1}J_{0}(\alpha\theta_{1}^{+})+\nonumber \\
 & V_{n-1}J_{0}(\alpha\theta_{1}^{-})\Big\}+\Big\{\gamma_{2}\vert V_{n}\vert^{2}+\gamma^{(0)}\vert U_{n}\vert^{2}\Big\} V_{n}-\label{eq:024}\\
 & i\sigma\Big\{\alpha V_{n}\Big[(U_{n+1}^{*}U_{n}-U_{n+1}U_{n}^{*})J_{1}(\alpha\theta_{2}^{+})-(U_{n-1}^{*}U_{n}-\nonumber \\
 & U_{n-1}U_{n}^{*})J_{1}(\alpha\theta_{2}^{-})\Big]+V_{n+1}J_{0}(\alpha\theta_{1}^{+})-V_{n-1}J_{0}(\alpha\theta_{1}^{-})\Big\}\nonumber \\
 & +\Omega\Big\{\alpha J_{1}(\alpha\theta_{0})(U_{n}\vert V_{n}\vert^{2}+U_{n}^{*}V_{n}^{2})+U_{n}J_{0}(\alpha\theta_{0})\Big\}.\nonumber
\end{align}
Furthermore, the pdes Eqs. \eqref{eq:023} and \eqref{eq:024} can
be written in terms of $iU_{n,t}=\delta H_{av}/\delta U_{n}^{*}$
and $iV_{n,t}=\delta H_{av}/\delta V_{n}^{*}$ respectively, where
the averaged Hamiltonian is:
\begin{equation}
\begin{aligned}H_{av}= & \sum_{n}\Big[-\varGamma J_{0}(\alpha\theta_{2}^{+})(U_{n+1}U_{n}^{*}+U_{n+1}^{*}U_{n})\\
 & -\varGamma J_{0}(\alpha\theta_{1}^{+})(V_{n+1}V_{n}^{*}+V_{n+1}^{*}V_{n})\\
 & +i\sigma J_{0}(\alpha\theta_{2}^{+})(U_{n+1}U_{n}^{*}-U_{n+1}^{*}U_{n})\\
 & -i\sigma J_{0}(\alpha\theta_{1}^{+})(V_{n+1}V_{n}^{*}-V_{n+1}^{*}V_{n})\\
 & +\Omega J_{0}(\alpha\theta_{0})(U_{n}V_{n}^{*}+U_{n}^{*}V_{n})\\
 & +\frac{1}{2}(\gamma_{1}\vert U_{n}\vert^{4}+\gamma_{2}\vert V_{n}\vert^{2})+\gamma^{(0)}\vert U_{n}\vert^{2}\vert V_{n}\vert^{2}\Big].
\end{aligned}
\label{eq:025}
\end{equation}
By comparing the averaged Hamiltonian in Eq.~(\ref{eq:025}) with the
original Hamiltonian Eq.~\eqref{eq:005}, their terms coincide if one
rescales the tunneling or hopping constant $\varGamma$, the SOC parameter
$\sigma$ and the Rabi frequency $\Omega$ according to Eq.~(\ref{eq:009}).
It should be noted here that the averaged equations only valid for
$t\leq1/\epsilon$ with accuracy of $\mathfrak{\mathcal{O}}\left(\epsilon\right)$
\cite{Sanders}. 

\end{document}